\documentclass[a4paper,11pt]{article}
\pdfoutput=1 % if your are submitting a pdflatex (i.e. if you have
\usepackage{amssymb,amsmath}

\usepackage{jcappub} % for details on the use of the package, please
\usepackage[T1]{fontenc} % if needed
\usepackage{newtxtext,newtxmath}
\usepackage{ae,aecompl}
\usepackage{hyperref}
\usepackage{cleveref}
\usepackage{changes}
\usepackage{enumitem}
\def\prd{Physical Review D}
\def\pra{Physical Review A}
\def\jcap{JCAP}

\def\apj{ApJ}

\def\apjs{ApJ Suppl. Ser.}

\def\aap{A\&A}

\def\araa{ARAA}

\def\nar{New Astron. Rev.}

\def\physrep{Phys. Rep.}

\def\nphysa{Nucl. Phys. A}

\def\beq#1{\begin{equation}\label{#1}}
\def\eeq{\end{equation}}
\def\beqa#1{\begin{eqnarray}\label{#1}}
\def\eeqa{\end{eqnarray}}

\def\myfrac#1#2{\left(\frac{#1}{#2}\right)}

\def\comment#1{\relax}

\newcommand{\be}{\begin{eqnarray}}
\newcommand{\ee}{\end{eqnarray}}

\title{\boldmath  X-ray signature of antistars in the Galaxy}

\author[a,e] {A.E.~Bondar,}
\author[b,c] {S.I.~Blinnikov,}
\author[d] {A.M.~Bykov,}
\author[e] {A.D.~Dolgov,}
\author[c,e,1] {K.A.~Postnov \note{Corresponding author.}}

\affiliation[a]{Budker INP, Lavrentieva 11, Novosibirsk, 630090, Russia}
\affiliation[b]{NRC ``Kurchatov institute'' - ITEP, B. Cheremushkinskaya 25, 117218, Moscow,  Russia }
\affiliation[c]{Sternberg Astronomical Institute, M.V. Lomonosov Moscow State University,\\ 13, Universitetskij pr., 119234, Moscow, Russia}
\affiliation[d]{Ioffe Institute, Politechnicheskaya 26, St Petersburg}
\affiliation[e]{Department of Physics, Novosibirsk State University, \\Pirogova 2, 630090, Novosibirsk, Russia}

\emailAdd{pk@sai.msu.ru}
\emailAdd{byk@astro.ioffe.ru}
\emailAdd{dolgov@fe.infn.it}
\emailAdd{sergei.blinnikov@itep.ru}

\abstract{
The existence of macroscopic objects from antimatter (antistars) is envisaged in some models of baryogenesis. Searches for antistars have been usually carried out in gamma-rays originated from hadronic annihilation of matter. In astrophysically plausible cases of the interaction of neutral atmospheres or winds from antistars with ionized interstellar gas, the formation of excited $p\bar p$ and He$\bar p$ atoms precedes
the hadronic annihilation. These atoms rapidly cascade down to low levels before annihilation giving rise to a series of narrow lines which can be associated with the hadronic annihilation gamma-ray emission. 
The most significant are L (3p-2p) 1.73 keV line (yield more than 90\%) from $p\bar p$ atoms, and  M (4-3) 4.86 keV (yield $\sim 60\%  $) and  L (3-2) 11.13 keV (yield about 25\%) lines from $^4$He$\bar p$ atoms. These lines can be searched for in dedicated observations by the forthcoming sensitive X-ray spectroscopic missions XRISM, \textit{Athena} and  \textit{Lynx} and in wide-field X-ray surveys like SRG/\textit{eROSITA} all-sky survey.
}

\begin{document}

\maketitle
\flushbottom
\section{Introduction \label{s-intro}}

According to the accepted conviction, the universe in our neighborhood consists solely of matter, and the existence of
macroscopically large antimatter objects is fully excluded. This is in accordance with the conventional picture
that the excess of baryons over antibaryons in the Universe is generated by Sakharov's mechanism \cite{1967JETPL...5...24S}.
%~\cite{ADS-BG}.
In models
with explicit C and CP violation, the baryon asymmetry
\be
\eta = (N_B - N_{\bar B} ) / N_\gamma \approx 6 \cdot 10^{-10}
\label{beta}
\ee
is a universal constant over the whole universe. Here $N_{B(\bar B)}$ are the number densities of baryons or antibaryons, and
$N_\gamma = 411/$cm$^3$ is the current number density of photons in the cosmic microwave background (CMB) radiation.

If the charge symmetry is broken spontaneously, then domains of matter and antimatter may exist but the size of domains
is expected to be cosmologically large, at the Gigaparsec level  \cite{1998ApJ...495..539C}.
Therefore, the accepted faith is that if even
macroscopically large objects or regions consisting of antimatter exist, they should be far away from us at the edge
of the universe or at least at distances beyond 10 Mpc \cite{1976ARA&A..14..339S}.
%~\cite{Steigman}.

On the other hand, about a quarter of a century ago, a physical mechanism was suggested leading to the possible formation of compact
antistars over normal matter-dominated background \cite{1993PhRvD..47.4244D, 2009NuPhB.807..229D}.
%~\cite{AD-JS,DKK}.
According to the proposed scenario, the baryogenesis
in the bulk of the universe proceeded in the usual way giving rise to the observed small baryon asymmetry (\ref{beta}),
while in a small by volume part of the universe, bubbles with a very high baryonic number  (called HBB) might be created.
The model allows for the formation of HBBs with both signs of the baryon asymmetry, positive (matter) and negative (antimatter). HBBs
with a size comparable to the cosmological horizon at the QCD (quantum chromodynamics) phase transition at $T\sim 100$~MeV  mostly give rise to primordial black holes (PBHs) which could make a sizeable fraction (if not all) of cosmological dark matter. Such PBHs could have masses in a wide range from a fraction to thousands of solar masses \cite{2016JCAP...11..036B,2020JCAP...07..063D}.
%or even  100\% of it.
Of course, there is no noticeable difference between a PBH and an anti-PBH.

Smaller HBBs could form
compact stars or antistars.  They would be created in the very early universe after the QCD  phase
transition at $T \sim 100 $ MeV. Probably, they can be now observed as peculiar stars in the Galaxy: too old stars, very fast-moving stars,
and stars with a highly unusual chemical content \cite{2004PThPh.112..971M,2005PhRvD..72l3505M,2020PhRvD.102b3503A}, for a review see \cite{2018PhyU...61..115D}.
%\cite{ad-ufn}.
They should also populate the galactic
halo. The fraction of these strange stars and/or antistars among the stellar population is model-dependent but may be quite high.

The possibility of the existence of a population of antistars in the Galaxy and observational limits on their density were analyzed
in several papers \cite{2007NuPhB.784..132B,2014JETPL..98..519D,2014PhRvD..89b1301D,2015PhRvD..92b3516B}.
%~\cite{cb-ad,AD-VAN-MIV,AD-SIB,SIB-AD-KAP}.
It was concluded that quite a noticeable fraction
of antistars relative to ordinary stars is not excluded by the present observations.  
As discussed in detail in ref. \cite{2015PhRvD..92b3516B}, the restrictive limits derived for the ''normal'' antistars from gamma-ray background (e.g. \cite{2014HyInt.228..100V})  are
not applicable for antistars produced in scenarios \cite{1993PhRvD..47.4244D}-\cite{2014PhRvD..89b1301D}, and such antimatter objects may abundantly populate the Galaxy without contradicting the gamma-ray background limits.

Recently, the idea that antistars may be our Galactic neighbors attracted new attention. In ref.
\cite{2020PhLB..80735574S},
%~\cite{anti-DM},
it was suggested
that cosmological dark matter could entirely consist of macroscopic antimatter objects and in ref.
\cite{2021PhRvD.103h3016D}
a possible indication to antistars in the Galaxy based on gamma-ray  \textit{Fermi}/LAT measurements was reported. Quoting the latter paper:
``We identify in the catalog 14 antistar candidates not associated with any objects belonging to established gamma-ray source classes and with a spectrum compatible with baryon-antibaryon annihilation.''

In this paper, we propose
a new signature tagging electromagnetic signal from annihilation in outer layers of antistars. The idea is to search for antistars in the Galaxy through X-rays in the  $\sim(1-10)$ keV energy band. This X-ray radiation is expected to be produced  in the process of the cascade transition
%to lower energy levels
from bound states of antiproton-proton (protonium) or antihelium-proton atoms before $p \bar p$-annihilation and must accompany the gamma-ray radiation from the annihilation.

Note that the \textit{Fermi} candidate antistars are not true stars but large regions in the sky plane (of order tens angular minutes in diameter corresponding to \textit{Fermi}/LAT gamma-ray error boxes) comprising a huge number of stars. Obviously, not only antistars but also objects consisting of ordinary matter can be sources of gamma-quanta with energies above 100 MeV. Therefore, the objects showing the X-ray signature proposed in our paper will lead to an unambiguous indication of their antimatter nature and much better localization if they are detected in future observations.

\section{Line emissions from protonium decays}

\begin{figure}
    \centering
    \includegraphics[width=\textwidth]{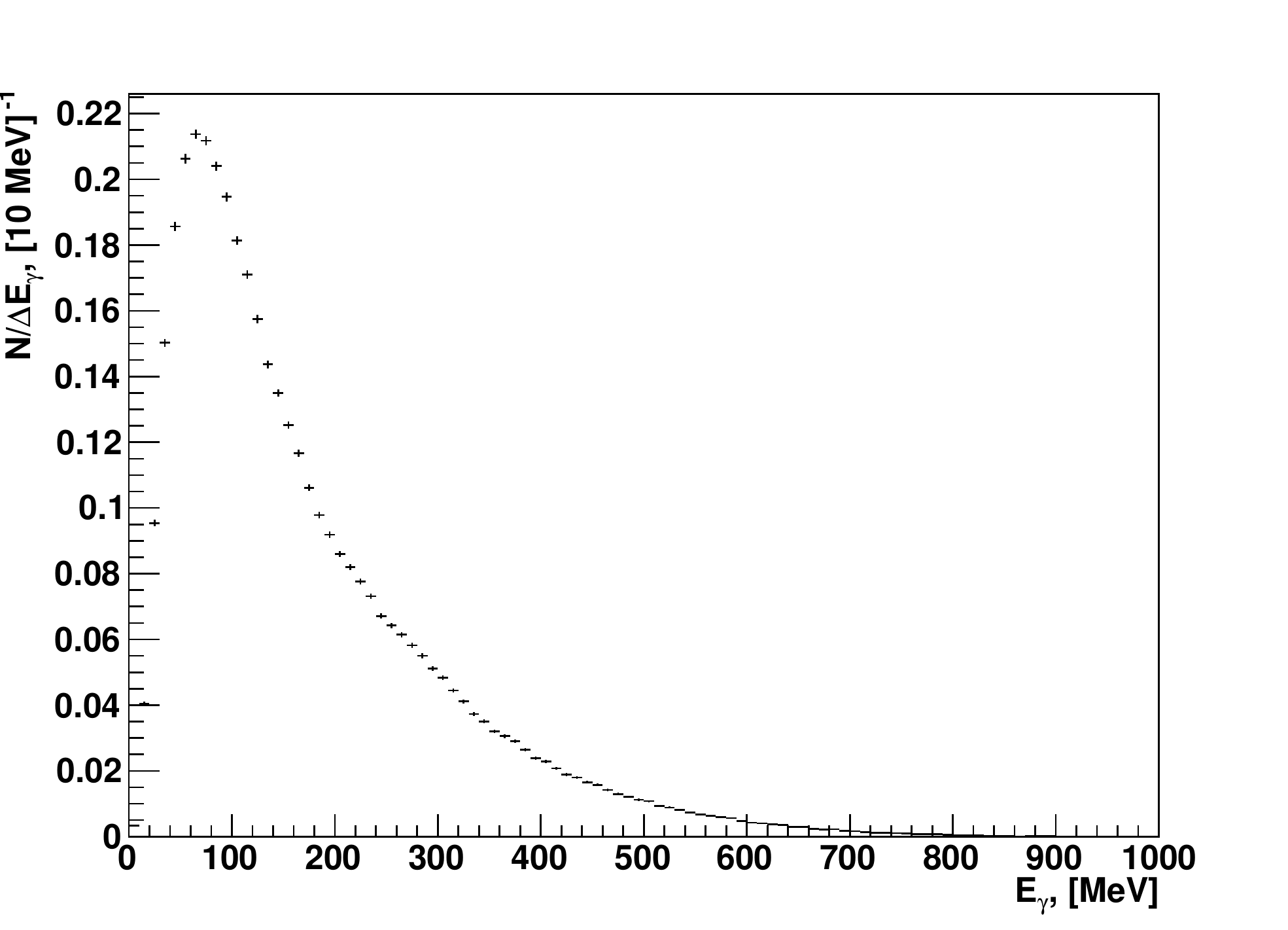}
    \caption{Gamma-ray spectrum from hadronic $p\bar H$ annihilation. The mean number of $\gamma$-quanta per event is $<N_\gamma>=4.12$. The mean number of $\gamma-$quanta above 100 MeV is 2.63 per event. The mean energy radiated in gamma-rays per one annihilation is $\Delta E_\gamma=617.5$~MeV. Calculations by Geant4 code.}
    \label{f:anncrosssec}
\end{figure}

Usually, the antimatter objects are being looked for through the
search for energetic photons with energies of hundreds MeV from $\pi^0$ decays into two $\gamma-$quanta,
with $\pi^0$ emerging from $p \bar p$ - annihilation.
The energy spectrum of such photons is shown in Fig.\ref{f:anncrosssec}. This spectrum was calculated by Geant4\footnote{{\tt version Geant4.10.06 (06.12.2019); https://geant4.web.cern.ch}}
\cite{Allison:2006ve} simulation of the low-energy antiproton annihilation in a small liquid hydrogen target.
The spectrum suggests that
the mean energy radiated in gamma-rays per one annihilation is $\Delta E_\gamma=617.5$~MeV. The mean number of $\gamma$-quanta per one annihilation is $\langle N_\gamma\rangle=4.12$ with the mean number of $\gamma-$quanta above 100 MeV is $\langle N_\gamma\rangle(>100 \mathrm{MeV})=2.63$ per event. Note that the target's material and structure do not significantly affect the annihilation spectrum which is appropriate for the considered problem.

However, before annihilation, protons and antiprotons could form atomic-type excited bound states (`protonium', Pn), similar to $e^+e^-$-positronium (Ps) atoms.
One could expect that in the process of de-excitation of Pn, an antistar could emit not only $\sim 100$-MeV gamma-rays but a noticeable flux of X-rays with energies in the keV range.

Let us briefly remind the basic physics of $p\bar p$ annihilation through a protonium state (see, e.g, \cite{2002PhR...368..119K} for a comprehensive review).
A protonium atom can form during the interaction of a $p$ with neutral (or molecular) antimatter. In the ionized matter-antimatter interaction, the annihilation mainly proceeds through the direct hadronic channel, the cross-section of the radiative recombination of a Pn atom formation from the $p\bar p$ interaction being $(m_e/m_p)^{3/2}\approx 10^{-5}$ times smaller (see \cite{1976ARA&A..14..339S}  for the matter-antimatter annihilation cross-sections).

A Pn atom forms when a proton (antiproton) interacts with an anti-hydrogen (hydrogen) atom, $\bar p+H\to (p,\bar p)+e^-$. The formation of a Pn atom effectively occurs when the energy of the proton in the lab frame is below the electron's ionization threshold in atomic hydrogen, $E=1$Ry$\simeq 13.6$ eV. At lower proton energies, the cross-section increases as $\sim 1/\sqrt{E}$ \cite{2005PhRvA..71e2717O}.
Therefore, in the present paper, we will mainly focus on astrophysical situations where the antistar's atmosphere (or its stellar wind) is neutral or interacts with the neutral interstellar medium (ISM).

The fraction of energy from the bolometric luminosity released in X-ray from the Pn cascades (the cascade yield),
$f_\mathrm{X}=L_\mathrm{X}(Pn)/L_a$, depends on the pressure \cite{1989NuPhA.503..885R}. However, an upper limit on the bolometric X-ray yield per unit cascade can be estimated as the binding energy of a Pn atom, $\approx 12.5$~keV (ignoring small QED corrections $\mathcal{O}(\alpha^2)$). Therefore, the maximum X-ray yield per one Pn atom is $R_{p\bar p}=(1/4) \alpha^2m_pc^2=12.5$ keV ($\alpha$ is the fine structure constant, $m_p$ is proton's mass). As the cascade time of a Pn atom is very short \cite{1989NuPhA.503..885R}, in rarefied media the X-ray luminosity in X-ray lines will be determined by the formation rate of Pn atoms, $\dot N_a$:
$L_\mathrm{X}\le 12.5\,\mathrm{keV}\times \dot N_a $, i.e. the maximum possible X-ray fraction in the Pn-cascade during matter-antimatter annihilation is $f_\mathrm{X,max}=\alpha^2/8\approx 6.7\times 10^{-6}$. 

The line energies for the K (Lyman) $2p\to 1s$, L (Balmer) $3 d \to  2p$ and M (Paschen) $4f\to 3d$ transitions in a Pn atom are 9409 eV, 1737 eV and 607 eV, respectively. Calculations of the X-ray line yields from the Pn decays produced in $p\bar H$ interactions at low densities \cite{1990PhRvA..41.3460C} suggest that most intensive X-ray lines should be Balmer (L) lines (up to 97\%), the 2p-states of the $p\bar p$ atom being rapidly annihilated.
The produced X-ray line width is determined by the $p\bar p$ annihilation from the 2p-state and is about $\Delta E\sim \mathcal{O}(0.1$~eV) (see the discussion and experimental measurements in  \cite{2002PhR...368..119K}). 
As the total energy released in gamma-rays during $p\bar H$ annihilation is about 617 MeV (see Fig. \ref{f:anncrosssec}), the expected fraction of X-ray line emission flux from one Pn annihilation is $f_\mathrm{X,H}\approx0.95\times (1.7\, \mathrm{keV}/617.5\,\mathrm{MeV})\approx 2.5\times 10^{-6}$, i.e. a factor of three lower than the simple estimate given above.

\section{keV X-ray emission from protonium cascades in Galactic antistar candidates\label{s-galactic}}

Presently, it is difficult to reliably assess the number and physical parameters of antistars in the Galaxy. Estimates in paper \cite{2020PhRvD.102b3503A} are model-dependent and based on the assumption that gamma-ray emission from hadronic annihilation occurs in the putative antistar candidates during Bondi-Hoyle-Littleton accretion of the interstellar gas.
It is straightforward to estimate the expected X-ray flux associated with the gamma-ray annihilation emission in the case where the $p\bar p$ annihilation was preceded by the protonium formation.

%The possible Pn creation in the Galactic antistars interacting with ISM discussed in the previous Sections suggest that 
A crude estimate of the X-ray flux emitted in 1.7-keV narrow line from the gamma-ray sources -- candidates to antistars found from \textit{Fermi} 10-year LAT catalogue in ref. \cite{2021PhRvD.103h3016D} is:
\beq{e:fermi}
  F_\mathrm{X}(1.7\,\mathrm{keV})\sim f_\mathrm{X,H}\times F_\gamma (0.1-100\,\mathrm{GeV})\approx  10^{-17}[\mathrm{erg\,cm^{-2}s^{-1}}] \myfrac{F_\gamma(0.1-100\mathrm{\,GeV})}{4\times 10^{-12}\mathrm{erg\,cm^{-2}s^{-1}}}
  %\myfrac{E_\mathrm{X}}{1.7\mathrm{\,keV}}\myfrac{200 \mathrm{MeV}}{E_\gamma}
\eeq
(see Table \ref{tab:sources_specs} for individual candidates).
This low flux is unlikely to be uncovered with the current instrumentation even with the stacked detection strategy.

The forthcoming JAXA/NASA XRISM (X-Ray Imaging and Spectroscopy Mission) telescope will be able to provide a 5-4 eV resolution in the 0.3-12 keV energy range \cite{2018_XRISM}. The expected effective area of the \textit{Resolve} spectrometer is $\sim 210$~cm$^2$ \footnote{{\tt https://xrism.isas.jaxa.jp/research/documents/index.html}}. For approximately $10^6$~s observations, only a few 1.7-keV photons from a source with an X-ray flux of $\sim 3\times 10^{-17}$~erg cm$^2$ s$^{-1}$ can be collected.
A more promising could be measurements from the ATHENA X-IFU X-ray spectrometer which is expected to have a spectral resolution of $\sim 2$~eV and an effective area of $\sim 8000$ cm$^2$ in the keV range \cite{2016SPIE.9905E..2FB,2020AN....341..224B}. 
The 1.7 keV Balmer line of protonium falls within the sensitivity limit of the advanced-technology Lynx X-ray Observatory \cite{2019JATIS...5b1001G}, which will be able to measure X-ray fluxes down to $\sim 10^{-19}$~erg cm$^2$ s$^{-1}$ with an angular resolution of better than $\sim 1''$. While the X-ray line fluxes of the individual source candidates estimated in Table \ref{tab:sources_specs} are below the expected sensitivity of the XRISM telescope one may try to search for antistars through stacked source detection in serendipitous surveys of the next generation X-ray observatories. In this case, the appropriate choice of the antistar candidates can be made from the unidentified \textit{Fermi} source sample.

Note also that the most prominent 1.7-keV X-ray line from Pn decay in antistars should not interfere with other X-ray emission lines expected from stars. As no real stellar antistar counterparts have been identified inside large \textit{Fermi} error boxes, it is reasonable to compare real \textit{Chandra} X-ray spectra from different stars. For example, observations of solar-type dwarfs show that at energies around 2 keV only MgXII ($\lambda = 8.421$ \AA $\approx 1.46$~keV) emission is visible in spectra of young star EK Dra, the older stars demonstrating no visible emissions \cite{2010PNAS..107.7158T}. A giant binary nearby (about 13 pc) star Capella  \cite{2007A&A...461..679R} used for the grating observations calibration \cite{2003ARA&A..41..291P} demonstrates a rich variety of coronal X-ray lines, with MgXII and SiXIV ($\lambda = 6.182$ \AA $\approx 1.99$~keV) around 2 keV. Therefore, the possible 1.7 keV emission from a solar-type dwarf antistar and even an evolved anti-giant is unlikely to interfere with stellar coronal X-ray lines from highly ionized species.

\begin{table*}
\centering
\caption{\label{tab:sources_specs}
Estimate of possible Pn X-ray flux $F_X$ from 4FGL \textit{Fermi} antistar candidates. Energy flux $F_\gamma$ (0.1-100) GeV from \cite{2021PhRvD.103h3016D}.}
\begin{tabular}{lcc}
\hline
\hline
 Name              & $F_\gamma$ (0.1 - 100 GeV) & $F_X$ (1.7 keV) \\
 &   erg cm$^{-2}$ s$^{-1}$ & erg cm$^{-2}$ s$^{-1}$  \\
\hline
 4FGL J0548.6+1200 & $(4.2\pm 0.9) \times 10^{-12}$ & $\sim 1\times 10^{-17}$ \\
 4FGL J0948.0-3859 & $(2.5\pm 0.7) \times 10^{-12}$ & $\sim 6.3\times 10^{-18}$  \\
 4FGL J1112.0+1021 & $(2.5\pm 0.5) \times 10^{-12}$ & $\sim 6.3\times 10^{-18}$  \\
 4FGL J1232.1+5953 & $(1.8\pm 0.3) \times 10^{-12}$ & $\sim 4.5\times 10^{-18}$ \\
 4FGL J1348.5-8700 & $(3.0\pm 0.6) \times 10^{-12}$ & $\sim 7.5\times 10^{-18}$ \\
 4FGL J1710.8+1135 & $(2.5\pm 0.6) \times 10^{-12}$ & $\sim 6.3\times 10^{-18}$ \\
 4FGL J1721.4+2529 & $(3.3\pm 0.5) \times 10^{-12}$ & $\sim 8.3\times 10^{-18}$  \\
 4FGL J1756.3+0236 & $(4.4\pm 1.0) \times 10^{-12}$ & $\sim 1.1\times 10^{-17}$ \\
 4FGL J1759.0-0107 & $(5.9\pm 1.3) \times 10^{-12}$ & $\sim 1.5\times 10^{-17}$\\
 4FGL J1806.2-1347 & $(9.4\pm 2.2) \times 10^{-12}$ & $\sim 2.4\times 10^{-17}$ \\
 4FGL J2029.1-3050 & $(2.6\pm 0.6) \times 10^{-12}$ & $\sim 6.5\times 10^{-18}$ \\
 4FGL J2047.5+4356 & $(1.4\pm 0.4) \times 10^{-11}$ & $\sim 3.5\times 10^{-17}$ \\
 4FGL J2237.6-5126 & $(2.3\pm 0.5) \times 10^{-12}$ & $\sim 5.8\times 10^{-18}$ \\
 4FGL J2330.5-2445 & $(1.6\pm 0.4) \times 10^{-12}$ & $\sim 4.0\times 10^{-18}$\\
 \hline
\end{tabular}
\end{table*}

\section{Associated $e^+e^-$ emission and optical lines}

During the formation of a Pn atom from the interaction of a proton with $\bar H$ in the upper atmosphere of an antistar, a slow positron $e^+$ is ejected. It should form a positronium $e^+e^-$ atom in the surrounding medium before annihilation. Since during the Pn cascade transitions and annihilation, on average, three gamma-quanta with energy $E_\gamma\approx 200$~MeV are created per 600 MeV \textbf{(see Fig. 1)}, the accompanying $e^+e^-$ annihilation flux from the \textit{Fermi} antistar candidates is expected to be:
\beq{e:e+e-}
\left.\frac{dN}{dAdt}\right|_\mathrm{e^+e^-}\approx 2\times\frac{1}{3}\left.\frac{dN}{dAdt}\right|_{\gamma}\approx  6\times10^{-9}\frac{\mathrm{ph}}{\mathrm{cm^2s}}
\myfrac{F_\gamma(0.1-100\mathrm{\,GeV})}{3\times 10^{-12}\mathrm{erg\,cm^{-2}s^{-1}}}\myfrac{200\mathrm{\,MeV}}{E_\gamma}\,.
\eeq
(The factor 2 above is due to at least two photons are emitted in the Ps annihilation).
This is much smaller than the current 511~keV detection capabilities \cite{2016A&A...586A..84S}. 

Other possible traces of the Pn atom cascade decays could be searched for in the optical-UV range. Indeed, a Pn bound state is expected to be formed highly excited, $n\gtrsim 30$ \cite{2002PhR...368..119K}.  The Pn energy levels are
\beq{}
E_n\approx -\frac{12.5}{n^2}\mathrm{keV}\approx -13.9 \,\mathrm{eV}\myfrac{30}{n}^2\,,
\eeq
and the cascade transitions $(n,l=n-1)\to (n-1,l-2)$ pass through the optical-UV range: $\Delta E_n\approx 0.93\,\mathrm{eV}\myfrac{30}{n}^3$. 
The spacing of the optical lines is $\Delta E/E=\Delta \lambda/\lambda=2/n$ is much larger than their expected Doppler width, which could be their distinctive feature.
Clearly, the number of quanta in an optical emission line from the \textit{Fermi} antistar candidates should be approximately equal to the number of Pn atoms, therefore
$
\left.\frac{dN}{dAdt}\right|_\mathrm{opt}\approx  \left.\frac{dN}{dAdt}\right|_{\gamma}\approx 3\times 10^{-9}\frac{\mathrm{ph}}{\mathrm{cm^2s}}\,.
$
%\myfrac{\Phi_\gamma(0.1-100\mathrm{\,GeV})}{ 10^{-10}\mathrm{ph\,cm^{-2}s^{-1}}}\,.

For a NIR line at $\lambda = 4\mu \mathrm{m}$ with Doppler width $\Delta\lambda/\lambda=3\times 10^{-6}$ this photon flux corresponds to m$_\mathrm{AB}\sim 24.3^\mathrm{m}$ magnitude. While well within the rich by modern telescopes (e.g., JWST\footnote{{\tt https://www.stsci.edu/jwst/science-planning/proposal-planning-toolbox/sensitivity-and-saturation-limits}}), this flux in emission line is too small to be distinguishable on the background of stellar photons (for example, a solar-type star at a distance of 10 kpc would have m$_{AB}(4\mu \mathrm{m})\approx 18^\mathrm{m}$ \cite{2018ApJS..236...47W}). Therefore, the measurement of such weak emission lines is challenging and requires a dedicated technique (for example, the use of stellar coronographs to shield the stellar light, e.g. like the coronograph to be installed on the Nancy Grace Roman Space Telescope\footnote{{\tt https://roman.gsfc.nasa.gov/coronagraph.html}}).
    
%too faint even for very large telescopes. 
 %\textbf{Should a possible stellar counterpart be found inside the large \textit{Fermi} candidate error boxes (for example, by the proposed X-ray line signatures), that faint optical lines could be searched for in dedicated measurements by large aperture telescopes.}

\section{Helium-antiproton and antihelium-proton cascade X-ray lines from antistars}

Enhanced production of helium and metals is an important feature of the non-standard nucleosynthesis in HBBs with high $\eta$ \cite{2004PThPh.112..971M,2005PhRvD..72l3505M,2020PhRvD.102b3503A}. Therefore, it is instructive to consider the X-ray lines arising from cascade decays of $^4\bar{\mathrm{He}}p$ and $^4$He${{\bar p}}$ atoms that can be created in the interaction of ISM with antistars. The possible AMS-02 detection of non-thermal antihelium-4 nuclei\footnote{S. Ting, Latest Results from the AMS Experiment on the International Space Station (2018)} 
may indicate the presence of antistars in the Galaxy because its secondary production in the universe is challenging (see, e.g., the discussion in \cite{2019PhRvD..99b3016P}). Note that the magnetically active antistars may accelerate antihelium nuclei in stellar flares.

In a $^4$He$\bar p$ atom, the hadronic annihilation starts dominating the cascade transitions already at the third level. The $^4$He $\bar p$ atom transitions series M (4-3) at 4.86 keV [yield (57 $\pm 3)\%$] and  L (3-2) 11.13 keV [yield (25 $\pm 5)\%$] \cite{1991ZPhyA.338..217S} are certainly of interest given that the forthcoming X-ray spectroscopic mission \textit{XRISM} \cite{2020arXiv200304962X} is sensitive in 0.3-12 keV energy range. 
A simple estimate of the fractional flux in these lines relative to the annihilation gamma-ray flux can be done in the same way as for $p\bar{p}$ decays presented in the end of Section 2:  $f_\mathrm{X,He}(4.86\mathrm{keV})\approx0.57\times (4.86\, \mathrm{keV}/617.5\,\mathrm{MeV})\approx 4.5\times 10^{-6}$, $f_\mathrm{X,He}(11.3\mathrm{keV})\approx0.25\times (11.3\, \mathrm{keV}/617.5\,\mathrm{MeV})\approx 4.6\times 10^{-6}$, i.e. about twice as high as for $p\bar{p}$ decay X-ray lines. These estimates depend on the assumed He abundance, and for the solar He abundance must be decreased accordingly. However, they are around the same as for $f_\mathrm{X,H}$.
These lines can be even more important from the observational viewpoint because the protonium $p\bar p$ 3d-2p transition line energy 1.73 keV is close to the detector's Si K-shell complex lines and, therefore, a high energy resolution is needed to distinguish the lines.
Note also that in a hot medium with ionized hydrogen, when protonium atoms will not be formed, even a small mixture of He$^+$ ions would give rise to the effective (with atomic cross-sections) formation of He$\bar p$ atoms. Then only $^4$He$\bar p$ cascade X-ray lines will be produced.
 
A dedicated search for $^4$He$\bar p$ 4.86 keV and 11.13 keV lines in all-sky surveys like \textit{Spectrum-RG} \cite{2021arXiv210413267S} could also constrain the collective contribution of X-ray emission from antistars with enhanced He abundance.  

Like in the case of antiprotonium, $^4$He$\bar p$ atoms are formed at excited levels with principal numbers $28<n<35$. After an antiproton is captured, the remaining electron is rapidly ejected due to the internal Auger effect, and the process of ion ($^4$He$\bar p$)$^+$ radiative deexcitation begins along the circular sequence from $n\sim 30$ \cite{1988PhLB..203....9R}. Neglecting the strong interaction shift, the energy between consequent levels is 
\beq{}
\Delta E_n\approx 3.72 \mathrm{eV} \myfrac{30}{n}^2\frac{2n-1}{(n-1)^2},
\eeq
and these lines can be probed in the optical range too. 

\section{Possible astrophysical sources}

As shown above, it is challenging to test the Pn formation before hadronic annihilation in the   \textit{Fermi} Galactic antistar candidates. However, there could be favorable astrophysical conditions for the Pn creation from the interaction of hypothetical antistars with ISM. 

Consider first an antistar with mass $M$ and the standard (solar) abundance residing in the Galactic ISM with density $\rho_0$ (or number density $n_0$) and moving with the velocity $v$. There can be distinct cases of the star's interaction with ISM depending on the evolutionary state of the antistar (main-sequence or evolved) and its space velocity.

If the star is on the main sequence and has an insignificant mass loss via stellar wind, the interaction with ISM will be characterized by the gravitationally
captured mass-rate (the Bondi-Hoyle-Littleton accretion). The relevant physical scale is the Bondi radius,
\begin{equation}
    \label{e:RB}
    R_B=\frac{2GM}{(v^2+c_s^2)}
\end{equation}
where $v$ and $c_s$ is the star's velocity and ISM sound speed, respectively, $G$ is the Newtonian gravitational constant. The Bondi mass accretion rate is
\begin{equation}
    \label{e:dotMB}
    \dot M_B=\pi R_B^2 \rho_0\sqrt{v^2+c_s^2}
\end{equation}

If a star has significant proper mass-loss rate $\dot M_w$, the interaction with ISM will be characterized by the radius $R_w$, where the dynamical pressure of the stellar wind  is balanced by the ISM pressure, $\rho_w v_w^2\sim \rho_0c_s^2$:
\begin{equation}
    \label{e:Rw}
    R_w=\sqrt{\frac{\dot M_w (v_w/c_s)}{4\pi \rho_0 c_s}}
\end{equation}
If $R_w>R_B$, the ISM accretion is insignificant and the stellar wind directly interacts with ISM. In our estimates, we will consider the standard galactic ISM with particle number density $n_0\sim 1$ cm$^{-1}$ (corresponding to a density of $\rho\sim 10^{-24}$ g cm$^{-3}$) and sound velocity $c_\mathrm{ISM}\sim 1$~km s$^{-1}$. The interaction with such an ISM depends on the stellar space velocity $v$ and stellar wind mass-loss rate $\dot M_w$. We will consider four different cases: an antistar with or without stellar wind and typical galactic disk dispersion velocity $v\sim 10$ km s$^{-1}$ or halo velocity $v\sim 10^{-3}c$ ($c$ is the speed of light), summarized in Table \ref{tab:table}. Note that in both cases stellar velocities are supersonic.

\begin{table}[]
    \centering
    \caption{Possible cases of interaction with ISM of an antistar with velocity $v$ and wind mass-loss rate $\dot M_w$ (see text). The last column shows expected X-ray flux from Pn cascade transitions from a Galactic antistar at 1 kpc. }
    \label{tab:table}
    \begin{tabular}{lcccc}
    \hline\hline
          Velocity   & Stellar wind & Gamma-ray annihilation & Pn cascade X-ray& X-ray flux, $d=1$ kpc\\
    of antistar&  $\dot M_w$ [$M_\odot$~yr$^{-1}$] & luminosity, $L_a [\mathrm{erg\, s^{-1}}]$ & luminosity, $L_X [\mathrm{erg\, s^{-1}}]$ & $F_X [\mathrm{erg\,cm^{-2} s^{-1}}]$\\
    \hline
     $10$ km s$^{-1}$ &  $\approx 0$  & $\sim 10^{32}$ & $\lesssim 2.5\times 10^{26}$ &$\lesssim 2.5\times 10^{-18}$\\
         10 km s$^{-1}$ &  $\approx 10^{-10}$ & $\sim 4\times 10^{36} \dot M_{w,-10}$ & $\lesssim  10^{31}\dot M_{w,-10}$ & $\lesssim  10^{-13}\dot M_{w,-10}$ \\
         $10^{-3} c $ & $\approx 0$ & $\sim 0$ & $\sim 0$\\
         $10^{-3} c $ & $\approx 10^{-10}$ & $\sim 4\times10^{36} \dot M_{w,-10}$ & $\lesssim  10^{31}\dot M_{w,-10}$ & $\lesssim  10^{-13}\dot M_{w,-10}$  \\
         \hline
    \end{tabular}
\end{table}

\begin{enumerate}
    \item Low-velocity  case, $v\sim 10$~km s$^{-1}$,
    weak stellar wind\footnote{Solar wind mass-loss rate is  $\sim 10^{-14} M_\odot$ per year.}, $\dot M_w\approx 0$.

    In this case, the Bondi radius is $R_B\approx 2\times 10^{14}[\mathrm{cm}] m v_6^{-2}\approx 2.7\times 10^3 R_\odot v_6^{-2}$ (here $m\equiv (M/M_\odot)$, $v_6\equiv v/(10^6\mathrm{cm\, s^{-1}})$, $R_\odot \approx 7\times 10^{10}$~cm is the solar radius). The BHL accretion rate from ISM onto such a star is $\dot M_B\approx 1.8\times 10^{11}[\mathrm{g\, s^{-1}}]m^2v_6^{-3}$. The total annihilation rate is thus $\dot N_a\sim 10^{35}\mathrm{s}^{-1}$, and the bolometric annihilation luminosity is \textbf{$L_a\sim 617\,\mathrm{MeV}\times \dot N_a\sim  10^{32}\mathrm{erg\,s^{-1}}$}. The maximum possible Pn-cascade X-ray luminosity is then
    $L_\mathrm{X}=f_{\mathrm{X,H}}\times L_a\approx 2.5\times 10^{26}[\mathrm{erg\, s^{-1}}]m^2v_6^{-3}$.

        \item Low-velocity case, $v\sim 10$~km s$^{-1}$, moderate stellar wind $\dot M_w\approx 10^{-10} M_\odot$~yr$^{-1}$. The wind terminal velocity is about the ISM sound speed, no strong shock is formed.

In this case, $R_w\approx 7 \times 10^{16}\mathrm{[cm]}\dot M_{w,-10}^{1/2}n_0^{-1/2}(v_w/c_s)^{1/2}\gg R_B$. Here $\dot M_{w,-10}\equiv \dot M_w/(10^{-10} M_\odot$~yr$^{-1})$. The cross-section of Pn formation under these conditions is about $10^{-16}$~cm$^2$ \cite{2005PhRvA..71e2717O}.
    The formation of Pn states occurs in a layer with a width of $\sim 1/(n_0\sigma)\sim 10^{16}\mathrm{cm} < R_w$. The Pn formation rate is $\dot N_a=\dot M_w/m_p\sim 4\times 10^{39}M_{w,-10} \,\mathrm{s^{-1}}$ suggesting  a bolometric annihilation gamma-ray luminosity of $L_a\sim 4\times 10^{36} [\mathrm{erg\, s^{-1}}]\dot M_{w,-10}$. The Pn-cascade X-ray luminosity is $L_\mathrm{X}\lesssim  10^{31}[\mathrm{erg\, s^{-1}}]\dot M_{w,-10}$. 

    \item High-velocity case, $v\sim 10^{-3}c$, weak stellar wind $\dot M_w\simeq 0$.

    This case is more relevant if antistars populate the Galactic halo and move with the halo virial velocities $v\sim 10^{-3} c$.    In this case, the Bondi radius is very small, $R_B\sim 4 R_\odot v_{-3}^{-2}$ (here $v_{-3}=10^{-3}(v/c)$), the BHL accretion is inefficient, $\dot M_B\sim 10^7 [\mathrm{g\, s^{-1}}]m^2n_0v_{-3}^{-3}$, the associated annihilation luminosity is very weak.

    \item High-velocity case, $v\sim 10^{-3}c$, moderate stellar wind, $\dot M_w\simeq 10^{-10} M_\odot$~yr$^{-1}$.

    If an antistar with stellar wind moves with the Galactic virial velocity through ISM, a strong bow shock is formed ahead of the star. The matter behind the shock front is fully ionized.
    The front shock radius is estimated from the dynamical wind pressure balance: $R_{sh}=\sqrt{\dot M v_w/(4\pi \rho_0 v^2)}\simeq 7 \times 10^{16}\mathrm{[cm]}\dot M_{-10}^{1/2}\rho_{-24}^{-1/2}v_{-3}^{-1/2}(v_w/v)^{1/2}$, where
    $\rho_{-24}=\rho/(10^{-24}\mathrm{g\,cm^{-3}})$ is the ISM density. For slow winds expected from red giants $v_w/v\sim 3\times 10^{-5}$, and the shock radius is $R_{sh}\sim 4\times 10^{14}\mathrm{[cm]}\dot M_{-10}^{1/2}\gg R_B$. 

    This case is more complicated than the interaction of a cold slow stellar wind with ISM. The energy of protons downstream the shock is $E_p=(3/16) m_pv^2\sim 170$~eV.
    $\sigma\sim 10^{-23}$~cm$^2$ \cite{1976ARA&A..14..339S}, and the protons penetrate much deeper into the stellar wind before annihilating through the hadronic channel.
    The cross-section of the Pn formation strongly depends on the proton's energy and below $\sim$~1 Ry~$=13.6$~eV sharply increases up to a few $\times 10^{-16}$~cm$^2$.
    
    However, the hot protons rapidly loose energy by interacting with neutral stellar wind. The proton stopping length in hydrogen is [M.J. Berger et al. 1993 ICRU Report 49]
    \begin{equation}
    \label{e:lp}
        l=\frac{E_p}{dE_p/dx}=\frac{E_p}{4\pi r_e^2 m_ec^2\beta_p^{-2}(\rho_w/m_p) L(\beta_p)}
    \end{equation}
    Here $\beta_p=v_p/c=3/4(v/c)\sim 10^{-3}$ is the proton's downstream thermal velocity, $L(\beta_p)$ is the proton's stopping number. For low proton  energies, it can be evaluated only from experiments, and is dominated by nuclear scatterings. For an estimate of the stopping length of slow protons, we can use the ionization energy loss of slow protons in electron gas $L(\beta_p)\approx \log(2m_ev^2/\hbar\omega_0)$, where $\omega_0\approx 5.6\times 10^4\sqrt{n_e}$~rad s$^{-1}$ is the electron plasma frequency \cite{Lindgard54}. For the assumed parameters, $L(\beta_p)\sim \mathcal{O}(10)$.
    
   It is convenient to recast equation (\ref{e:lp}) in the form
   $$
   l=\frac{1}{n_0\sigma_T L(\beta_p)}\myfrac{1}{3}\myfrac{m_p}{m_e}
   \myfrac{v_w}{v}^2\beta_p^4
   $$ 
   ($\sigma_T=(8\pi/3) r_e^2$ is the Thomson cross section). Then   
  $l\lesssim 10^{11}\mathrm{[cm]} (\beta_p/10^{-3})^2(v_w/10\, \mathrm{km\, s}^{-1})^2$, which is much smaller than the shock front radius $R_{sh}\sim 4\times 10^{14}$~cm. Therefore, the ISM protons are expected to form Pn atoms inside the wind in analogy with the low-velocity case considered above. 
\end{enumerate}

If a typical antistar has an atmosphere like that of our Sun, then modelling by Kurucz's ATLAS code\footnote{http://kurucz.harvard.edu/programs.html} shows that the bulk of proton annihilations takes place in essentially neutral plasma. Then the data on antiproton annihilations in laboratory neutral media is quite applicable for proton annihilations in Solar-like atmospheres of antistars. Different possible atmospheric structures of antistars or their coronas deserve a special investigation.

Also note that at advanced evolutionary stages, antistars can experience a huge mass-loss, $\dot M_w\sim 10^{-6} M_\odot$ yr$^{-1}$, for a short time (e.g., at the asymptotic giant branch). In this case, as seen from Table \ref{tab:table}, the annihilation luminosity can reach $10^{40}$ erg s$^{-1}$, comparable to the gamma-ray emission of the entire galaxy. Presently, this is definitely not the case in our Galaxy, but if antistars are sufficiently abundant, there can appear a bright extragalactic gamma-ray and X-ray source. Future sensitive instruments can probe this possibility.

\section{Constraints from electron-positron  annihilations}

The interaction of an antistar with ISM should be necessarily accompanied by electron-positron annihilation. Therefore, the existing observations of the $e^+e^-$ Galactic emission should be taken into account when constraining the possible Galactic antistar populations.

The characteristic 511 keV annihilation radiation from the Milky Way was discovered about 50 years ago in balloon-borne experiments. Since then it has been measured by many space missions (for a review, see \citep{2011RvMP...83.1001P,2020NewAR..9001548C}). The recent analysis of {\sl INTEGRAL SPI} Ge detector data \citep{2016A&A...586A..84S} has confirmed the presence of the main extended 511 keV components with a total significance of about 58 $\sigma$. The 511 keV fluxes estimations depend on the models of the extended Galactic emission. The total 511 keV flux from the Milky Way was estimated to be  $(2.74 \pm 0.25) \times 10^{-3}$ ph cm$^{-2}$ s$^{-1}$. The bulge component contributed about $(0.96 \pm 0.07) \times 10^{-3}$ ph cm$^{-2}$, and the 511 keV flux ratio of the bulge to disk components was derived to be $(0.58 \pm 0.13)$. 

The total 511 keV Galactic disk emission suggests an annihilation flux of $F_\mathrm{e^+e^-}\sim 2\times 10^{-3}$ cm$^{-2}$ s$^{-1}$. 
For a fiducial distance of 10 kpc this rate corresponds to a total
positron production $\dot N_\mathrm{e^+e^-}\sim 10^{43}$~s$^{-1}$. If all positrons were formed in Pn 
$p\bar p$ annihilations in a purely hydrogen matter, the upper limit on the protonium production rate would be $\dot N_\mathrm{Pn}\sim  10^{43}$~s$^{-1}$, corresponding to  $\sim 2\times 10^{-7} M_\odot$~yr$^{-1}$. Therefore, an antistar with higher wind mass-loss rate is excluded by the $e^+e^-$ Galactic emission. 

Recently, ref. \cite{2021arXiv210903791S} reported an improved \textit{INTEGRAL} upper limit on the 511-keV flux from the Galactic satellite dwarf galaxy Ret II of 
$F_{511}<1.7\times 10^{-4}$ ph cm$^{-2}$ s$^{-1}$.
Given a distance of 30 kpc to this galaxy, this flux is translated to an $e^+e^-$ annihilation rate of about $\dot N_\mathrm{e^+e^-}\sim 10^{43}$ s$^{-1}$, comparable to the detected Galactic value.  The gamma-ray flux in the GeV range measured by Fermi-LAT from Ret II galaxy is inconclusive \cite{2020JCAP...02..012H}. The authors \cite{2021arXiv210903791S} do not exclude the 511 keV flux variability on a decade-long time-scale. In the frame of the astrophysical models discussed above (i.e., the interaction 
of antistars with ISM), the $e^+e^-$-flux variability might be due to the intrinsically variable stellar wind mass-loss rate.  

\section{Conclusions}

Antistars in the universe can be created in 
the modified Affleck-Dine baryogenesis mechanism \cite{1993PhRvD..47.4244D, 2009NuPhB.807..229D}. 
Presently, they could be observed as old halo stars with unusual chemical composition. 

In the present paper, we explored the possibility that the interaction of antistars with ISM gas can proceed with the formation of excited protonium atoms which rapidly cascade down before hadronic annihilation from 2p-states.  
The formation of Pn atoms most effectively occurs during the interaction of protons with neutral (or molecular) antimatter. This can happen if an antistar has a noticeable wind mass loss.
We considered the case where, in analogy with ordinary stars, antistars in the mass range from $\sim 0.8$ to $8 M_\odot$ at the end of their core nuclear burning can have cold low-velocity stellar winds  $\sim 10^{-8}-10^{-5} M_\odot$~yr$^{-1}$ \cite{1995A&A...297..727B,2008A&A...487..645R} (AGB antistars). 
The interaction of cold stellar winds from antistars with ISM can proceed through the formation of excited protonium $(p\bar p$) atoms. The protonium atoms cascade to the 2p-state producing mostly L (Balmer) 3d-2p X-rays around $\sim 1.7$~keV before the $p\bar p$ hadronic annihilation. 

Antistars formed in HBBs should have an enhanced helium abundance. Therefore, the 4.86 keV M (4-3) and 11.13 L (3-2) narrow X-ray lines from cascade transitions in $^4$He$\bar p$ atoms can also be associated with gamma-rays from hadronic annihilations. These lines are interesting from the observational point of view because
the protonium 3d-2p transition line energy 1.73 keV is close to the Si K-shell complex lines, which could hamper disentangling it from the background.

The expected Pn-decay X-ray line flux from individual Galactic \textit{Fermi} antistar candidates reported in \cite{2021PhRvD.103h3016D} is estimated to be much lower than the sensitivity of current X-ray detectors. However, searches for the cascade X-ray transitions from Pn and $^4$He$\bar p$ atoms associated with hadronic  annihilation gamma-rays can be exciting targets for the forthcoming X-ray spectroscopic missions like
\textit{XRISM}, \textit{Athena} and \textit{\textit{Lynx}}.
Sensitive wide-field X-ray surveys can also help to constrain the collective contribution of antistars in the universe.

\section*{Acknowledgement}
The authors thank the anonymous referee for careful reading of the paper and suggestions for improving the presentation of the results.
We are grateful to Andrey Sokolov for help in simulating the gamma-ray spectrum from hadronic $p\bar H$ annihilation.
The work of AD was supported by the RSF Grant 19-42-02004. 
SB acknowledges the support of RSF Grant 19-12-00229 in his work on stellar atmospheres. The work of KAP is supported by the Ministry of Science and Higher Education of the Russian Federation under contract 075-15-2020-778 (astrophysical appearance of antistars).

%\bibliographystyle{JHEP}
%\bibliography{antistars} % if your bibtex file is called example.bib

\providecommand{\href}[2]{#2}\begingroup\raggedright\endgroup

\end{document}